# HYBRIST Mobility Model - A Novel Hybrid Mobility Model for VANET Simulations

Wiseborn Manfe Danquah
Dept. of Computer Engineering
Istanbul Technical University
Istanbul-Turkey

D. Turgay Altilar
Dept. of Computer Engineering
Istanbul Technical University
Istanbul-Turkey

## ABSTRACT
Simulations play a vital role in implementing, testing and validating proposed algorithms and protocols in VANET. Mobility model, defined as the movement pattern of vehicles, is one of the main factors that contribute towards the efficient implementation of VANET algorithms and protocols. Using near reality mobility models ensure that accurate results are obtained from simulations.

Mobility models that have been proposed and used to implement and test VANET protocols and algorithms are either the urban mobility model or highway mobility model.

Algorithms and protocols implemented using urban or highway mobility models may not produce accurate results in hybrid mobility models without enhancement due to the vast differences in mobility patterns. It is on this score the Hybrist, a novel hybrid mobility model is proposed.

The realistic mobility pattern trace file of the proposed Hybrist hybrid mobility model can be imported to VANET simulators such as Veins and network simulators such as ns2 and Qualnet to simulate VANET algorithms and protocols.

## 1. INTRODUCTION
Vehicular Ad-hoc networks (VANET), the communication between vehicles and road side units (RSU) has attracted a lot of interest from the research community due to both the opportunity and challenges it presents.

From the late 90's, after the US Federal Communications Commission allotted the 75MHz of spectrum at 5.9GHz (from 5.850 to 5.925GHz) of the Dedicated Short Range Communications (DSRC)[1] for vehicle-to-vehicle and vehicle-to-infrastructure communications, a lot of research groups and projects have sprung up across universities and research institutions in America, Europe and Asia.

One of the main challenges in the full deployment of VANET is the testing of proposed protocols, standards and algorithms. Due to the unique characteristics of VANET such as extremely high speed of nodes, testing of protocols and standards using existing general network simulators may not produce accurate results.

Using real world test-beds are also extremely expensive and resource demanding. The best way for testing and evaluating algorithms and protocols therefore is by enhancing existing network simulations by incorporating realistic mobility models.

The high mobility of nodes in VANET makes mobility model selection one of the most important parameters when evaluating any protocol [2]. Mobility models determine the location of nodes in the topology at any given instant, which strongly affects network connectivity and throughput [3]. For results of simulations in VANET to be accurate, the mobility model should therefore be realistic.

Also results from simulations show that protocol performance may vary drastically across mobility models and performance rankings of protocols may vary with the mobility models used [4].

In VANET, the movement pattern of nodes or mobility model has been classified as urban mobility model and highway mobility model.

Most existing mobility models developed by the researchers deal with the vehicular movement within a city area [8]. The urban or city mobility model is characterized mostly by heavy node density (traffic) with nodes having averagely slow speed and a lot of intersections along roads. Urban mobility models have a lot of road side units (RSU) which can also be used in routing of communication data.

Different urban mobility models have been developed for VANET simulations. Most of these urban mobility models mainly deal with node movements at road intersections. They include the Random Way Point (RWM) model, Manhattan mobility mode, Rice University Model (RUM), Stop Sign (SSM), and Probabilistic Traffic Sign (PTSM), Traffic Light (TLM).

In the Random Waypoint Mobility model (RWM) [5] nodes are initially distributed randomly in the network. The velocity of the nodes is uniformly distributed with the minimum velocity of 0m/s. Nodes at initial position randomly selects final destination and moves towards in a given time. Upon reaching the destination the nodes wait for a period of time called the pause time. When the pause time expire the nodes selects a new random position, with random velocity and moves towards it. The movement of nodes continues using the procedure till the simulation ends.

There are a lot of derivatives of the random waypoint mode. Movement pattern of a mobile node using the Random Waypoint Mobility Model is similar to the Random Walk Mobility Model if pause time is zero and [minspeed, maxspeed] = [speedmin, speedmax] as can be seen in [4]. The Random Waypoint is the most used mobility model for VANET simulations

The Manhattan Mobility Model [6] uses grid road topology with equal-length square blocks within each grid. Vehicular nodes move in horizontal or vertical direction. At intersections of grids vehicles a probabilistic approach is adopted to determine the direction vehicles can move. The probability of nodes turning right or left is 0.25 whiles moving straight is 0.5. The city mobility model is a derivative of the Manhattan mobility model which also uses grids but movements at intersections is not based on a probabilistic approach.

The Stop Sign Model (SSM) [3] is an urban mobility model that uses stop signs as the traffic control mechanism at intersections. Every intersection of a street has a stop sign. Vehicles moving towards intersections stop at a signal for a specific time. The motion of vehicles is constrained by those in front. Therefore vehicular nodes cannot overtake nodes ahead of them unless it is a multi-lane street. Waiting of vehicular nodes at a stop sign create queues at intersections. The SSM may not be realistic as all intersections of streets cannot be equipped with stop signs as proposed.

The Probabilistic Traffic sign Model (PTSM) [3] is an improvement on the SSM. In the PTSM the stop signal in SSM is replaced by traffic signal at intersections. Vehicles stop at red traffic signals and move through intersections at green signals.

In modeling the PTSM a probabilistic scheme is used to approximate the operations of the traffic signals. Vehicular nodes at intersections with an empty queue, stop at the signal with a probability p and crosses the signal with a probability $(1 - p)$. When nodes decide to wait, the amount of wait time is randomly chosen between 0 and w seconds. The PTSM is more realistic than the SSM as it models the behavior of traffic lights and also prevents excessive wait times at intersections.

The Traffic Light Model [3] is the most realistic intersection mobility model. It avoids the heavy approximative approach of the PTSM and SSM by allowing the coordination of vehicles at intersections. Vehicles in a single-pair opposing traffic move through intersection at a green traffic light signal whiles the other pairs stop at a red signal. The free turn rule is applied for vehicles that need to turn at intersections.

In highway mobility model, nodes have extremely high speed and there exist little or no road traffic as well as road side units (RSUs). Highway Mobility Models (HWM) for VANET has not been fully explored; researchers who propose algorithms using highway mobility models do not explicitly state the details of the mobility patterns used for their simulations. [7][8]. One of the main research concerns in VANET is to find a model for highway mobility outside the city [10].

In view of this and in order to evaluate the proposed Hybrist hybrid mobility model, a simple Highway Mobility Model (HWM) was modeled using a real road map and road conditions of a simple highway in Istanbul.

## 2. MOTIVATION
Despite the different mobility models that have been developed for VANET simulations there are still unique mobility pattern that have not been modeled. Most mobility models proposed for VANET simulations may not suite the road conditions in all countries. Also due to the fact that protocol performance may vary drastically across mobility models and performance rankings of protocols may vary with the mobility models used [4], other mobility patterns should be explored to obtain realistic results in VANET simulations.

One of the reasons for proposing a novel Hybrid Mobility Model (HMM) is that, in many practical scenarios, multiple-models may exist within the same road network, due to the heterogeneity of nodes and users [9]. Hybrid mobility models, where a road exhibits both urban mobility pattern and highway mobility pattern at all instance of a road have not been studied.

This mobility pattern mostly exist in road architectures where there exist a dedicated Bus Rapid Transit system with a dedicated or restricted lane reserved for high speed buses surrounded by adjacent low speed heavy traffic lanes or roads.

The dedicated road have less traffic and extremely high speed nodes that are similar to highway mobility pattern while other lanes of the road have heavily dense traffic with slow nodes-urban mobility pattern. The picture taken in Istanbul covering multiple lanes including Metrobus lanes (located in the middle) shown in Figure 1 indicates the need of a Hybrid Mobility model for vehicular communication.

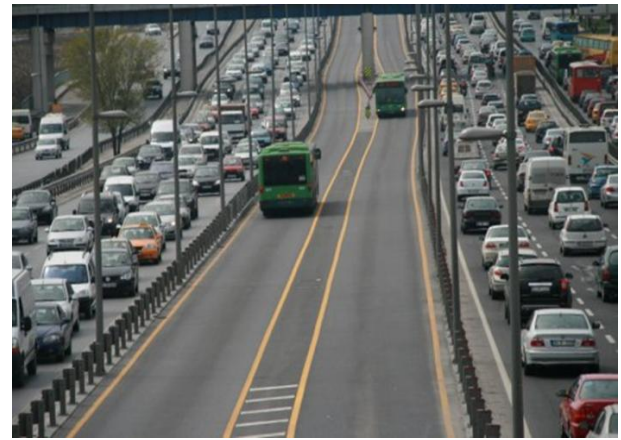

**Figure. 1: Sample of image exhibiting hybrid mobility pattern**

A Hybrid Mobility Model can be designed considering a) the middle lanes reserved for high speed nodes that always experience little or no traffic and b)the other multiple unrestricted lanes that experience heavy traffic with various slower speed nodes.

A lot of major important Bus Rapid Transit (BRT) [16] roads in some of the cities of Asia, Europe and the South America have such roads with the characteristics of the hybrid mobility pattern [17] The Istanbul Bus Rapid Transit (BRT) system, the first intercontinental bus rapid system [7] and other Bus Rapid Transit systems such as the TransMilenio (BRT system in Colombia)[26], and the TransJakarta in Jakata, Indonesia[25], have a lot of socio-economic importance.

The Transjakarta (TJ) records about 400 thousand passengers per day and it is expected to rise to about 800 thousand passengers per day by the year 2014 [18]. Despite its importance, a lot of road accidents and deaths occur on these roads. From January to July 2010 there were 237 accidents involving TransJakarta buses, resulting in 57 injuries and eight deaths [17]. Due to the importance as wells as the large number of accidents that occur on these BRTs, effective and efficient protocols and standards must be developed to suit these Hybrid Mobility Models (HMM) before the full deployment of VANET.

To design and propose algorithms and protocols for a vehicular traffic that includes BRT(s) a novel HMM must be designed and incorporated in VANET and network simulators.

In this paper, a novel HMM is proposed utilizing the Istanbul BRT system. This exhibits both properties of highway mobility and urban mobility at almost all instances of the road.

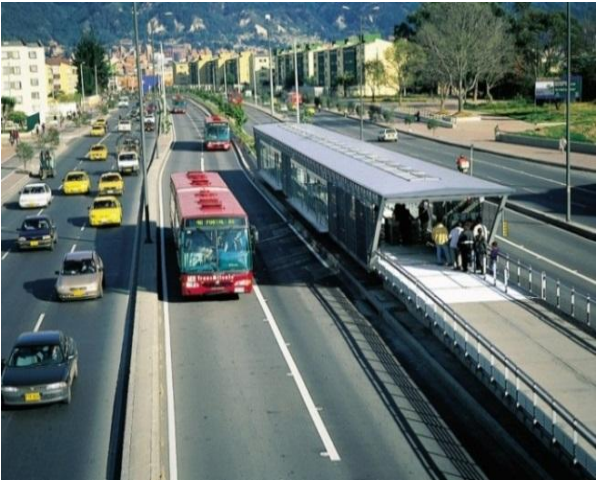

**Figure. 2: The TransMilenio (BRT system in Colombia) [26]**

## 3. RELATED WORK

Even though a lot of work has been carried out in MANET mobility modeling, VANET has not received much needed attention. One of the reasons can be due to the assertion by [12] that the study of realistic vehicular mobility models and their deployment is a challenging task.

Despite the fact that no similar work on modeling a hybrid mobility model (a combination of Urban and Highway mobility Models) for VANET has been proposed, some works have been carried out with the approach used in his paper, that is, the use of real world road map to design mobility models for VANET simulations.

In [3] the authors used real street maps extracted from the US censor bureau TIGER database [13] to model their proposed urban mobility models (Stop Sign (SSM), and Probabilistic Traffic Sign (PTSM), Traffic Light (TLM).) explained in the introduction section. Nidhi et al [14] used real world map extract from Google Earth and other GIS tools to generate mobility model to evaluate the performance of VANET in realistic environments.

The Street Random Way point (STRAW) [15] mobility model, was also proposed using real map data. The STRAW mobility model provides a functionality that simplifies traffic congestion by controlling vehicular mobility. It is however considered not realistic.

The term Hybrid Mobility has been used by a number authors to represent mobility models in VANET but in all these instances they combined two or more urban mobility models. Rajini et al [10] proposed hybrid mobility for clustering in VANET, but the mobility model was just the combination of two urban mobility models, the random waypoint and group mobility model.

## 4. THE ISTANBUL BRT SYSTEM

The Istanbul BRT system, a bus rapid transit (BRT) system is about 42km long and runs across the Istanbul strait which divides the continent of Asia and Europe. This makes the Istanbul BRT the first continental bus rapid transit [7].

The BRT system is a hybrid between a metro train and a metro bus system. It consists of a fully restricted dedicated right-of –way one lane (referred to as metro bus lane) for high speed metro buses, security patrol vehicles, ambulances and other emergency vehicles. This high speed restricted commercial metro bus lane has currently 33 bus stop stations with no intersections (closed bus transit system) and it is still growing. It carries approximately one million passengers in a day [3].

The restricted lane (metrobus lane) is surrounded by two unrestricted opposite one-way direction three lane roads with less speed and heavy traffic vehicular nodes called main roads. There may other roads which flows in parallel to previously mentioned two with less speed and relatively low traffic which is called side roads ('yanyol' in Turkish). Since side roads appear in parallel with just some sections the main roads, focus was made on only the relation between Metrobus lane and main roads within the context of this paper.

The mobility pattern spans between the towns of Avcılar, Zincirlikuyu and the end of the Bosporus Bridge and Söğütlüçeşme.

The nature of the mobility pattern of the Istanbul BRT system presents a unique vehicular mobility model for the development and implementation of new VANET algorithms. Already designed algorithms may need further enhancement or modification to achieve better results before implementation on the Istanbul BRT and other BRTs.

## 5. MODELLING AND SIMULATIONS

In evaluating the difference in terms of the performance of the proposed Hybrid Mobility Model (HMM), Urban Mobility Model (UMM) and Highway mobility (HWM), were compared using the results obtained from simulations.

To generate realistic mobility models for simulations, the mobility pattern of a section (block) of the Istanbul BRT was observed. Inserted below is an image of a section of the Istanbul BRT system showing the metrobus lane and main roads.

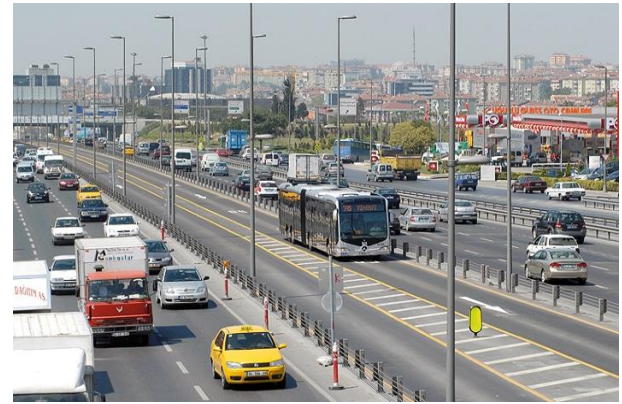

**Figure 3: A section of the Istanbul BRT system showing the main roads and BRT [4]**

The section (block) was extracted from the online map, OpenStreetmap tool. The extracted mobility pattern was then edited using the map editing tools Java OpenStreet map editor, JOSM [22] and e-World [23].

Roads were modeled by connected line segments (edges). Each edge has attributes such as road identifier, the speed limit of lanes, number of lanes and priority of lanes. The modeled road network was imported to SUMO and the required traffic was generated. The Figure 4 shows a section of the road network of the Istanbul BRT exported from e-World into SUMO

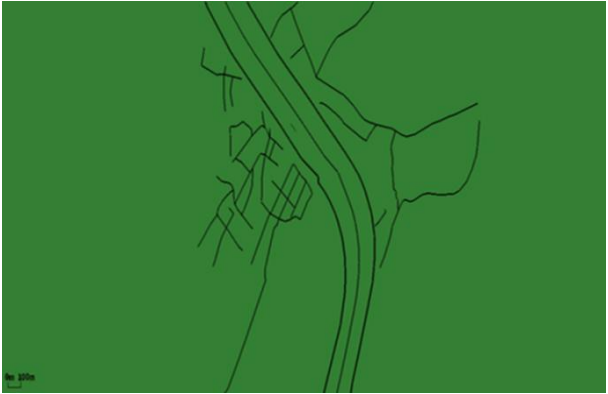

**Figure. 4: Istanbul BRT road network model in SUMO**

The DROUTER tool based on the Dijkstra's algorithm [24] is used to compute the routes of nodes. The Car-following model [11] implementation in SUMO, is employed in modeling the vehicular traffic. The map file (.net.xml) and corresponding route files (.rou.xml) were used to generate sumo trace files (.sumo.tr) which is then imported to MOVE and converted to ns2 format for network analysis.

The sumo.tr file can also be visualized in sumo-gui as shown in Figure 5.For the Urban Mobility the Stop Sig mobility model (SSMM) and traffic light model were incorporated along the section on the streets.

The number of stop stations for the metrobus lane, main road and the highway mobility models are shown in table 1 below. Other parameters such as the speed limit of lanes, the number of lanes used in modeling of the various mobility models in SUMO are also presented. The parameters were obtained from the highway authority in Istanbul Turkey which corresponds to the values obtained from the OpenStreet map.

**Table 1: Parameters used in modeling**

| Road type | No. lanes | Speed of lanes (m/s) | No. of stop stations |
|-----------|-----------|----------------------|----------------------|
| Metrobus-lane | 1 | 33.33 | 5 |
| Main road | 3 | 13.89 | 10-15 |
| HWM | 4 | 33.3 | 0-5 |
| UMM | 4 | 13.89 | 10-20 |

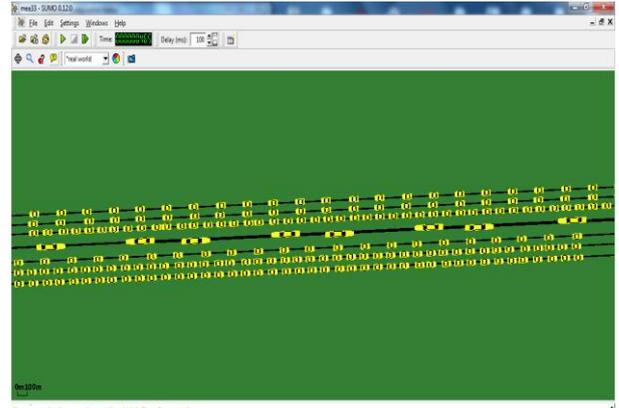

**Figure. 5: Screen shot of SUMO simulation**.

# 6. NETWORK SIMULATIONS

Using ns-2.35 CBR generation tool with seed value of 2 and transmission interval of 0.05micro seconds, varying CBR sources were generated for the different mobility models. Several instances for each CBR value were run for each mobility model. Parameters used for simulations have been presented in the Table 2.

**Table 2: Table of simulation parameters**

| Parameter | Value |
|-----------|-------|
| Simulation time | 1000 seconds |
| Simulation area | 6380m X 1934m |
| Mobility Models | Hybrist Mobility Model(HMM), UMM, HWM |
| Number of Vehicles | 160, 200,250 |
| Routing Protocol | AODV |
| Transmission Range | 50m,100m,150m,200m 250m and 300m |
| CBR Sources | 5-20 |
| Packet size | 512 Bytes |
| Routing Protocol | AODV |
| Transmission Range | 50,100,150,200,250m |
| Bandwidth | 10 Mbps |
| MAC Protocol | 802.11p |

The screenshot of simulation in ns-2 utilizing nam trace file is shown in Figure.6. The node movement and communication among of vehicular on all the lanes of the road have been captured in this figure.

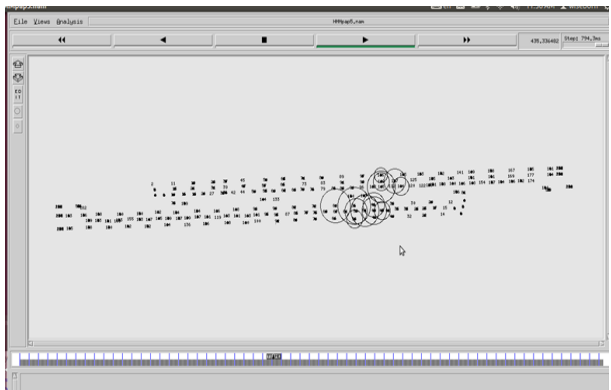

**Figure. 6: Screenshot of simulation in ns-2 nam interface of HMM simulation**

# 7. PERFORMANCE ANALYSIS

The results of performance metrics that were compared to determine the uniqueness of the Hybrist, hybrid mobility model (HMM) in relation to the highway mobility model (HWM) and urban mobility model (UMM) proposed are discussed in details in this section.

## 7.1. Packet Delivery Fraction (PDF)

Packet delivery ration (PDF), is the ratio of the number of successfully delivered data packets to all destination nodes and the number of data packet generated by all source nodes.

The PDF values from the three mobility models are used to determine which mobility model has the highest percentage of packets reaching their intended destination during the entire duration of the simulation. It can be observed from Figure 7 that the values of the HMM and UMM are higher than HWM. This could be due to the relatively higher node density, relatively lower speed of nodes thereby forming more stable link connections among nodes as compared to the HWM.

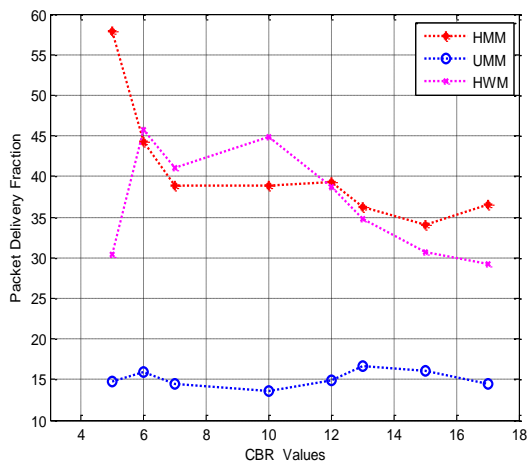

**Figure. 7: PDF vs. CBR sources**

## 7.2. End-End delay vs. CBR sources

The End-to-end delay can be defined as the time taken for a packet to be transmitted across a network from source to destination.

It is observed that the end to end values rise as the CBR sources increases as seen in the Figure 8. In the VANET environment as the CBR sources increase the number of packets contending for the same wireless channel increases. This leads to a lot of packet collision within the channel.

Large packet drop being the consequences of collision within the channel leads to large end to end delay values. The higher values and large rise in value for the UMM could attributed to the larger number of nodes than that of the HWM.

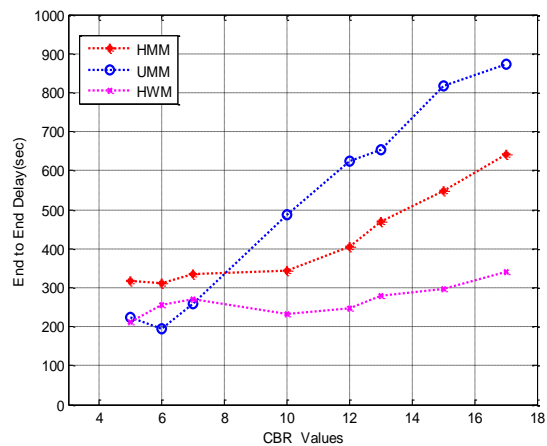

**Figure. 8: End-to-end delay vs. CBR sources**

## 7.3. Packet Loss

The packet lost value is the total number of packets dropped during simulation. The packet loss is mainly due to the instability nature of links between nodes. The Highway mobility (HWM) has nodes which are sparsely apart. This accounts for the large values of packet loss.

In Figure 9, Packet Loss vs. CBR sources HMM has slightly better performance than the UMM this can be attributed to the presence of extremely high nodes on the fast speed lane do not exist in the UMM. These nodes present extra faster link for data transfer when other links within the network break down. It ensures the fast transmission of data when already established links within the main road break down.

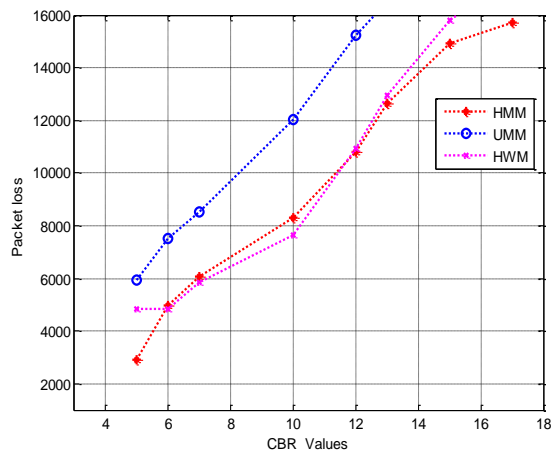

**Figure.9: Packet Loss vs. CBR sources**

## 7.4. Packet Delivery Ratio vs. Transmission Range

The effect of varying transmission range on the packet delivery ration is a good measure to determine how unique mobility models are.

From the graph of packet delivery ratio vs. varying transmission range shown in Figure 10, it could be observed that the packet delivery ratio decreases when the transmission range of nodes is increased. The HWM experiences huge packet loss even though the transmission range is increased because the signal strength of the nodes decreases with increasing transmission range. The HMM however performs better than the UMM as the mobility pattern remains relatively stable during peak and off-peak hours. The UMM mobility pattern may change drastically during off-peak hours.

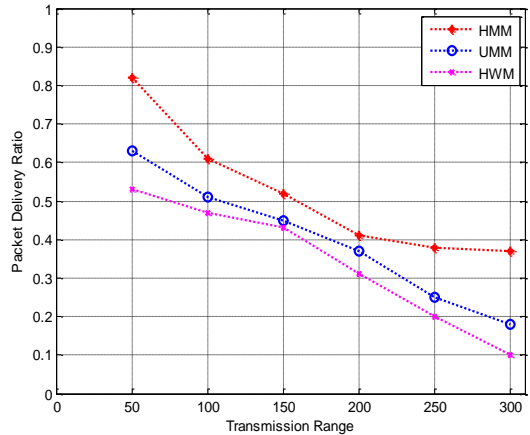

**Figure 10: Packet Delivery Ratio vs. Transmission range**

## 8. CONCLUSION

This paper presents a proposed Hybrid Mobility Model (HMM), the Hybrist for VANET simulations. Analysis of simulation results shows that, the Hybrid Mobility varies from both the Urban Mobility and simple Highway Mobility (HWM) in terms of performance metrics shown.

The use of real data from OpenStreet map makes the Hybrist mobility model more realistic for the design testing of VANET protocols before implementation. The Hybrist (HMM) presents the basis for further studies into the mobility models that are similar to both the Urban Mobility Model and Simple Highway Mobility Models in terms of speed of nodes and the node density in VANET environments.

Future works would implement more efficient routing protocols as well as examine the performance of already existing routing protocols (AODV, DSDV, DSR etc.) and other protocols in the Hybrid Mobility Model.

Evaluation of performance of Intelligent lane reservation systems: Nishkam R. et al (2007) [28], proposed an intelligent lane reservation system for highways. In their position paper, they proposed a model where drivers on highways are allowed to reserve a slot on a reserved high-priority lane. The high-priority lane has higher speed limit and less traffic than the other lanes. During the peak hours, where the highway may experience heavy traffic, the high-priority lane is only used by drivers who have made earlier reservation. Their proposed model, considered as a traffic congestion control scheme can be evaluated using the proposed HYBRIST- Mobility Model as they have the same characteristics.

Further studies would also be undertaken to determine level of details that should be considered for a near to realistic simulation of the proposed Hybrid Mobility Model. Mobility trace files of the Hybrid Mobility Model (HMM) May be downloaded from www.bb.itu.edu.tr.